\documentclass[12pt]{article}

\usepackage{amsmath}
\usepackage{amssymb}
\usepackage{amsfonts}
\usepackage{lscape}

\begin{document}

\fontsize{12}{6mm}\selectfont
\setlength{\baselineskip}{2em}

$~$\\[.35in]
\newcommand{\dss}{\displaystyle}
\newcommand{\raro}{\rightarrow}
\newcommand{\be}{\begin{equation}}

\def\sech{\mbox{\rm sech}}
\def\sn{\mbox{\rm sn}}
\def\dn{\mbox{\rm dn}}
\thispagestyle{empty}

\begin{center}
{\Large\bf The Einstein-Hilbert Action, }  \\    [2mm]
{\Large\bf Horizons and Connections With  }  \\    [2mm]
{\Large\bf Thermodynamics}   \\   [2mm]
\end{center}

\vspace{1cm}
\begin{center}
{\bf Paul Bracken}                        \\
{\bf Department of Mathematics,} \\
{\bf University of Texas,} \\
{\bf Edinburg, TX  }  \\
{78541-2999}
\end{center}

\vspace{3cm}
\begin{abstract}
It is shown that the Einstein-Hilbert action can be
constructed by minimizing free energy. The entropy used
to determine the free energy is determined on the horizon
of a black hole. Some further considerations with regard to
generalizations of these ideas to other situations of
physical importance are presented as well.
\end{abstract}

\vspace{2mm}
PACS: 04.20.Cv, 04.20.Fy, 04.60.Bc, 04.70.Dy
\vspace{2mm}

\newpage
\begin{center}
{\bf 1. Introduction.}
\end{center}

For some time, it has been known that there are deep connections
between the areas of gravity and thermodynamics {\bf [1-5]}. One of the
main reasons for this is that the principle of equivalence has
led naturally to the idea that gravity is a manifestation of
curved space-time {\bf [6-7]}. Gravity then is of geometric origin and can
be termed a geometric effect. In general relativity, the local 
inertial frames at the various events are patched together
in the presence of gravity to generate a curved space-time.
If the global space-time were flat, the different local
inertial frames would fit together to form an extended inertial
frame. The idea of gravity in general relativity as an aspect
of inertia merges with the elimination of the absolute set of
extended inertial frames. Since these distortions of space-time
can be regarded as surfaces or types of membranes which are
capable of being deformed, distorted or even wrapped up,
it is frequently the case that information from one region
is not accessible to observers in a different region.
Sakharov thought of gravitation as a metric elasticity. This
metric elasticity was based upon a microscopic structure
analogous to molecular structure behind material elasticity.
In this picture, any static space-time with horizon, space-time
dynamics can be regarded as similar to the thermodynamic
limit in solid state.
It is frequently the case that information from one region is not
accessible to observers in a different region. It is also
known that there is a deep relationship between entropy and
information, or accessibility of information. All told, it 
must be concluded that there must be many connections between 
gravity and thermodynamics, even just thinking at the 
classical level {\bf [8-9]}. There has also been a lot of
interest in looking at thermodynamic systems as extremal
hypersurfaces {\bf [10]}.

If the equivalence principle leads to the formulation of a
geometrical description of gravity, then there is the
possibility that membranes will arise in viewing space-time under various
circumstances. Regarding space-time as a membrane means
they can act as one-way interfaces for the
transfer of information, as discussed extensively by Padmanabhan {\bf [11-13]}. 
Thus, a connection is established
with thermodynamics, or a thermodynamic view, and a
particular instance in which such a surface-like structure,
or horizon, can physically appear is in the vicinity of a
black hole, namely the event horizon. Moreover, its possible 
horizons play a similar role in other relativistic situations 
as well. Thus the main objective here is to examine this
relationship, as well as to discuss
other physical situations in which similar ideas may play an important role.
In the case of a black hole, the information content entangled
across a horizon is proportional to the area of the horizon.
This brings with it the idea that the fundamental constant
characterizing gravity is the quantum of area $4 {\cal A}_P$,
which can contain about one bit of information. The situation 
is similar to the case of bulk matter made of individual atoms.
The conventional gravitational constant $G = {\cal A}_P c^3 / \hbar$
will diverge if $\hbar \raro 0$ when ${\cal A}_P$ is kept constant. 
In a way similar to the fact that  an atomic system does not
exist in this limit, space-time and gravity are inherently
quantum mechanical.

A new coordinate system can
be constructed by first transforming to the local inertial
frame and then using standard transformations between the
inertial coordinates and Rindler coordinates. In this context, 
it can be postulated that the horizon in the local Rindler
frame has an entropy per unit transverse area and then demand
that any feature of gravity have this incorporated in it. 
Of course, the static situation in terms of a 
black hole space-time corresponds
to an equilibrium state in the thermodynamic picture. Moreover, it
will be shown that by minimizing the free energy, this is
sufficient to lead to the correct Einstein-Hilbert action
principle for gravity. By using the local Rindler frame
and demanding that gravity must incorporate these thermodynamical
aspects leads in a natural way to the action functional itself.
The action is built up from its surface behaviour, 
which is to say that gravity is fundamentally holographic 
in nature. Finally, in addition to finding that Einstein's
equations are equivalent to the principle of minimization
of the free energy, some speculation as to the relevance
of these ideas to other systems in which the equivalence
principle plays a role, such as accelerating systems or
free fall, will be discussed at the end.

\begin{center}
{\bf 2. Gravitational Functional.}
\end{center}

It will be important to have some information concerning
the gravitational functional at the start. It is this functional
which will result from thermodynamic considerations.
Begin by considering some general properties of the
gravitational functional. The conventional action principle
for general relativity is the Einstein-Hilbert action,
which is given as {\bf [6,9]}
$$
S_{EH} = \frac{1}{16 \pi} \, \int R \sqrt{-g} \, d^4 x,
\eqno(2.1)
$$
where $R$ is the Ricci scalar curvature calculated using the 
components of the metric $g_{ij}$. In fact, (2.1)
can be reexpressed by making use of the following relation
$$
R \sqrt{-g} = \frac{1}{4} \sqrt{-g} M^{abcijk} \,
g_{ab,c} g_{ij,k} - \partial_j P^j = \sqrt{-g} {\cal L}_{quad}
- \partial_j P^j.
\eqno(2.2)
$$
In (2.2), comma indicates partial differentiation with respect to
the indicated coordinate. 
The factor needed to define ${\cal L}_{quad}$ is given by
$$
M^{abcijk} = g^{ck} ( g^{ab} g^{ij} - g^{ai} g^{bj} )
+ 2 g^{cj} ( g^{ai} g^{bk} - g^{ki} g^{ba} ),
\eqno(2.3)
$$
and $P^j$ is given by
$$
P^j = \sqrt{-g} \, g_{ac,i} ( g^{ac} g^{ji} - g^{ia} g^{cj} )
\equiv \sqrt{-g} V^j.
\eqno(2.4)
$$
The equality between the first term on the left and last on the
right in (2.2) is well known {\bf [14]}. If the Lagrangian of the
gravitational field is taken as the noninvariant quantity
$$
{\cal L}_{quad} = g^{ab} ( \Gamma^s_{a r} \Gamma^r_{b s}
- \Gamma^s_{a b} \Gamma^r_{r s}),
\eqno(2.5)
$$
then $\sqrt{-g} {\cal L}_{quad}$ differs from $\sqrt{-g} R$
by a divergence term of the form
$
{\partial( \sqrt{-g} V^j)}/{\partial x^j}.
$
From (2.2), ${\cal L}_{quad}$ can be identified to have the form,
$$
{\cal L}_{quad} = \frac{1}{4} \, M^{abcijk} \, g_{ab,c} g_{ij,k}.
$$
Upon differentiating both sides of this with respect to
$g_{uv,w}$, the following result is obtained
$$
4 \frac{\partial {\cal L}_{quad}}{\partial g_{uv,w}}
= M^{uvwijk} g_{ij,k} + M^{abcuvw} g_{ab,c}
$$
$$
= g^{wk} ( g^{uv} g^{ij} - g^{ui} g^{vj} ) \, g_{ij,k}
+ 2 g^{wj} ( g^{ui} g^{vk} - g^{ki} g^{vu} ) g_{ij,k}
$$
$$
+ g^{kw} ( g^{ij} g^{uv} - g^{iu} g^{jv}) \, g_{ij,k}
+ 2 g^{kv} ( g^{iu} g^{jw} - g^{wu} g^{ji}) \, g_{ij,k}
$$
$$
= 4 g_{ij,k} ( g^{ij} g^{kw} - g^{wj} g^{ik}).
\eqno(2.6)
$$
The first important result that is obtained from (2.2) and (2.6)
is a direct relationship between $P^j$ and Lagrangian ${\cal L}_{quad}$,
$$
g_{ab} \frac{\partial {\cal L}_{quad}}{\partial g_{ab,q}}
= g_{ij,k} ( g^{ij} g^{qk} - g^{ik} g^{qj}) - V^q =
\frac{1}{\sqrt{-g}} P^q.
\eqno(2.7)
$$
By replacing $P^j$ in (2.2), this result shows that the
scalar curvature can be put in the form,
$$
R = {\cal L}_{quad} - \frac{1}{\sqrt{-g}} \partial_k
[ \sqrt{-g} \, g_{ij} \frac{\partial {\cal L}_{quad}}
{\partial g_{ij,k}} ].
\eqno(2.8)
$$
Choose a coordinate system in which the metric has the form,
$$
ds^2 = g_{nn} \, (dx^n)^2 + g_{ij}^{\perp} \, dx^i dx^j,
\eqno(2.9)
$$
and $n=0, \cdots, 3$. For each choice of $n$, $i$ and $j$
run over the other three coordinates, assuming cross terms
vanish. Then $P^n$ in this coordinate system can be worked
out by using (2.4),
$$
P^n =- \frac{1}{\sqrt{-g}} \partial_k (g g^{kn} )
=- \frac{2}{\sqrt{g_{nn}}} \partial_n \sqrt{g^{\perp}}.
\eqno(2.10)
$$
With $C$ a constant, the normal to the surface $x^n =C$ is given by
$n^a = g_{nn}^{-1/2} \delta_n^a$, and the trace of the
extrinsic curvature of the $x^n=C$ surface is
given by
$$
K =- \nabla_a n^a =- \frac{1}{\sqrt{g_{\perp}}} \frac{1}{g_{nn}}
\partial_n \sqrt{g^{\perp}}.
$$
Consequently,
$$
\int_{V} \, d^4 x \partial_a P^a = \sum_{\partial V}
2 \int K \sqrt{g^{\perp}} \, d^3 x.
\eqno(2.11)
$$
The sum in (2.11) is over the boundary surfaces, so the total
divergence term can be expressed as the sum over the integrals
of the extrinsic curvatures on each boundary.
Comparing (2.11) with (2.8), a dynamical interpretation of
$K$ can be obtained
$$
2 K = n_c g_{ab} \frac{\partial {\cal L}_{quad}}{\partial g_{ab,c}}
= n_c g_{ab} \pi^{abc}.
\eqno(2.12)
$$
The quantity $\Pi^{ab} = n_c \pi^{abc}$ is the energy-momentum
conjugate to $g_{ab}$ with respect to the surface defined by the
normal $n_c$.

If the Lagrangian is taken to have the form ${\cal L} (q_A,
\partial_i q_A)$ and depends on a collection of dynamical
variables $q_A$, where in turn $A$ can denote a collection
of indices, a second Lagrangian can be obtained as follows
$$
{\cal L}_{\pi} = {\cal L} - \partial_i [ q_A
\frac{\partial {\cal L}}{\partial ( \partial_i q_A)} ]
= {\cal L} - \partial_i ( q_A p^{Ai}).
\eqno(2.13)
$$
For gravity $q_A$ would be taken as $g_{ab}$ with $A$
denoting a pair of indices. Each of these Lagrangians will 
lead to the same equations of motion, provided that $q_A$
is fixed while varying ${\cal L}$ and the $p^{Ai}$ are
fixed while varying ${\cal L}_{\pi}$. In the context of
gravity, Lagrangian ${\cal L}$ corresponds to the quadratic
Lagrangian while ${\cal L}_{\pi}$ corresponds to the
Einstein-Hilbert Lagrangian.

Consider any Lagrangian ${\cal L} ( q_A, \partial_i q_A )$
which contains dynamical variables $q_A$ such that the
Euler-Lagrange function will have the form
$$
E^A \equiv \frac{\partial {\cal L}}{\partial q_A} 
- \partial_i [ \frac{\partial {\cal L}}{\partial (\partial_i q_A)} ].
\eqno(2.14)
$$
Contracting this with $q_A$, (2.14) can be put in the form,
$$
q_A E^A = q_A \frac{\partial {\cal L}}{\partial q_A}
- \partial_i [ q_A \frac{\partial {\cal L}}{\partial (\partial_i q_A)} ]
+ ( \partial_i q_A ) \frac{\partial {\cal L}}{\partial ( \partial_i q_A)}.
\eqno(2.15)
$$
For example, suppose ${\cal L}$ is a homogeneous function of
degree $\mu$ in the $q_A$ and homogeneous of degree $\lambda$ in the
$\partial_i q_A$. Then (2.15) reduces to 
$$
q_A E^A = ( \lambda + \mu) {\cal L} - \partial_i [ q_A
\frac{\partial {\cal L}}{\partial ( \partial_i q_A)} ].
\eqno(2.16)
$$
In the case of gravity, $E^A$ works out to be
$$
E^A =- ( R^{ab} - \frac{1}{2} g^{ab} R) \sqrt{-g}.
\eqno(2.17)
$$
Contracting (2.17) with $q_A = g_{ab}$, it turns into
$$
q_A E^A = g_{ab} [ - (R^{ab} - \frac{1}{2} g^{ab} R)
\sqrt{-g} ] = R \sqrt{-g}.
$$
In this case, it can be verified that $\mu =-1$ 
and $\lambda =2$, so (2.16) becomes identical to (2.8).

\begin{center}
{\bf 3. Entropy and Horizons.}
\end{center}

To work entirely in the Lorentzian space-time, some
semi-classical features must be adopted. First, the time integration
must be restricted to a suitable finite range in defining the
action. Second, there should be a suitable surface term
in the action describing gravitational dynamics which acquires
a contribution from the horizon. The horizon is the only surface
or interface which is common to both the inside and outside regions.
Quantum entanglement effects across a horizon can only appear 
as a surface term in the action. The equivalence principle then
leads to the conclusion that the action functional describing 
gravity must contain certain boundary terms which can encode 
information which is equivalent to that present beyond the
horizon. The idea here is to determine this surface term from
general principles. This can be used to make a correspondence 
with entropy and out of the equations from 
thermodynamics, obtain the form of the full action for gravity.

In order to provide a local Lagrangian description, the sought
after boundary term has to be expressible as an integral of a
four-divergence. This allows the action functional to have the 
form,
$$
S_{grav} = \int \, d^4 x \, \sqrt{-g} {\cal L}_{grav} 
= \int \, d^4 x \, \sqrt{-g} ( {\cal L}_{bulk} +
\nabla_i W^i ) = S_{bulk} + S_{sur}.
\eqno(3.1)
$$
In (3.1) ${\cal L}_{bulk}$ is quadratic in the first derivatives
of the metric and of course
$$
\nabla_i W^i = (-g)^{-1/2} \, \partial_i 
[ (-g)^{1/2} W^i ],
\eqno(3.2)
$$
irrespective of whether $W^i$ is a general four vector or not.
It is required to determine $S_{sur}$ so that a connection with
entropy can be established. 

Let $(M, g_{ab})$ be a globally hyperbolic space-time, which
can be foliated by Cauchy surfaces, $\Sigma_t$, parametrized by 
a global time function, $t$. Let $u^a$ be the unit normal vector
field to the hypersurfaces $\Sigma_t$. The space-time metric
$g_{ab}$ induces a three-dimensional metric $h_{ab}$ on each
$\Sigma_t$, as discussed below. Let $t^a$ be a vector field on
$M$ satisfying $t^a \nabla_a t =1$, which is decomposed into its 
parts normal and tangential to $\Sigma_t$, by defining the
lapse function, $N$, and the shift vector, $N^a$, with respect
to $t^a$ by
$$
N =- t^a n_a = ( n^a \nabla_a t)^{-1},
\quad
N_a = h_{ab} t^b.
$$
The lapse function and shift vector are not considered dynamical,
since they merely prescribe how to move forward in time, a gauge transformation.
The horizon for a class of observers
arises in a specific gauge and $S_{sur}$ will in general depend
on the gauge variables $N$, $N_{\alpha}$. The lapse function $N$
plays a more important role than the $N_{\alpha}$, and so we
set $N_{\alpha}=0$ without loss of generality.

Next a $(1+3)$ foliation with the standard notion for the metric
components $g_{00} =-N^2$, $g_{0 \alpha} = N_{\alpha}$. Let
$u^i = (N^{-1},0,0,0)$ be the four-velocity of observers
corresponding to this foliation; that is, the normal to this
foliation. Let $a^i = u^j \nabla_j u^i$ be the related acceleration
and $K_{ij} =- \nabla_i \, u_j -u_i a_j$ be the extrinsic
curvature of the foliation, with $K \equiv K^i_i =- \nabla_i u^i$,
hence $K_{ij} u^i = K_{ij} u^j =0$ and $K_{ij}$ is purely spatial.

Consider a series of hypersurfaces $\Sigma$ with normals $u^i$.
The following differential geometric identity will be required {\bf [6]}.
Beginning with
$$
R_{ijka} u^a = ( \nabla_i \nabla_j - \nabla_j \nabla_i ) u_k,
$$
the following required identity is obtained,
$$
R_{ij} i^i u^j = g^{ac} R_{aicj} u^i u^j = u^i \nabla_a \nabla_i u^a
- u^j \nabla_j \nabla_a u^a
$$
$$
= \nabla_a ( u^j \nabla_j u^a) - ( \nabla_a u^i)(\nabla_i u^a)
- \nabla_j (u^j \nabla_a u^a) + ( \nabla_j u^j)^2
$$
$$
= \nabla_i ( K u^i + a^i) - K_{ij} K^{ij} + K^i_i K^j_j,
\eqno(3.3)
$$
where $K_{ij} = K_{ji} =- \nabla_i u_j - u_i a_j$ is the extrinsic
curvature with $K= K^i_i =- \nabla_i u^i$ and $K_{ij} K^{ij} =
( \nabla_i u^j)(\nabla_j u^i)$. In any space-time, there is the
geometric identity (3.3). In static space-time with $K_{ij}=0$,
this reduces to
$$
\nabla_i a^i = R_{ij} u^i u^j.
\eqno(3.4)
$$
Thus all possible vector fields $W^i$ which can be used in (3.1)
may be accounted for. It must be made up of $u^i$, $g_{ij}$ and
$\nabla_i$ acting only once, since the equations of motion must
be no higher order than two. Given these conditions, there is
only one vector field, $u^i$ itself, and only three vectors
$( u^j \nabla_j u^i, u^j \nabla^i u_j, u^i \nabla^j u_j)$ which
are linear in $\nabla_i$. The first one is the acceleration
$a^i = u^j \nabla_j u^i$. The second identically vanishes since
$u^j$ has unit norm and the third can be written as
$-u^i K$. Consequently, $W^i$ appearing in the surface term 
has to be a linear combination of the terms $u^i$, $u^i K$ at
lowest order. Hence $S_{sur}$ must have the form
$$
S_{sur} = \int \, d^4 x \sqrt{-g} \nabla_i W^i
= \int \, d^4 x \sqrt{-g} \, \nabla_i
[ \lambda_0 u^i + \lambda_1 K u^i + \lambda_2 a^i ],
\eqno(3.5)
$$
where $\lambda_j$ are numerical constants.

Let the region of integration be a four-volume $V$ bounded by
two space-like surfaces $\Sigma_1$, $\Sigma_2$ and two time-like
surfaces ${\cal S}$ and ${\cal S}_1$. The space-like surfaces 
are constant time slices with normals $u^i$, and the time-like
surfaces have normals $n^i$ such that $n_i u^i=0$.
The induced metric on the space-like surface
$\Sigma$ is $h_{ij} = g_{ij} + u_i u_j$, while the
induced metric on the time-like surface ${\cal S}$ is
$\gamma_{ij} = g_{ij} - n_i n_j$. These two surfaces intersect 
on a two-dimensional surface ${\cal Q}$ with induced metric
$\sigma_{ij} = h_{ij} - n_i n_j= g_{ij} - u_i u_j - n_in_j$.
In this instance, the first two terms in (3.5) contribute
only on $\Sigma_1$, $\Sigma_2$ with $t$ constant, while the third
term contributes on ${\cal S}$, that is on a horizon.
Consequently, on the horizon
$$
S_{sur} = \lambda_2 \, \int \, d^4 x \sqrt{-g} \nabla_i a^i
= \lambda_2 \int_{{\cal S}} \, dt d^2 x N \sqrt{\sigma}
( n_i a^i).
\eqno(3.6)
$$
In any static space-time with horizon, the integration over
$t$ becomes multiplication by $\beta = 2 \pi / \kappa$, where
$\kappa$ is the surface gravity of the horizon, since there
is a periodicity in the Euclidean sector. As ${\cal S}$ approaches
the horizon, $N (a_i n^i)$  in the integrand tends to $-\kappa$,
which is constant over the horizon. Thus on the horizon,
$$
S_{sur} =- \lambda_2 \kappa \int_0^{\beta} \, dt
\int d^2 x \sqrt{\sigma} = - 2 \pi \lambda_2 {\cal A}_{H},
\eqno(3.7)
$$
where ${\cal A}_H$ is the area of the horizon.

Treating the action as analogous to entropy, the information
blocked by a horizon, and encoded in the surface term, must
be proportional to the area of the horizon. Taking into
consideration non-compact horizons like Rindler, the entropy
or information content per unit area of the horizon is a
constant related to $\lambda_2$. 
Writing $\lambda_2 =-1/ 8 \pi {\cal A}_P$, where ${\cal A}_P$ 
is a fundamental constant with dimensions of area, the entropy 
associated with the horizon is
$$
S_H = \frac{{\cal A}_H}{4 {\cal A}_P}.
\eqno(3.8)
$$

\begin{center}
{\bf 4. Einstein's Equations Based on a Thermodynamic Argument.}
\end{center}

The information content which is entangled across
a horizon surface is proportional to the area of the horizon.
Consequently, there is a fundamental constant characterizing
gravity, the quantum of area. Consider a four-dimensional region
of space-time defined as follows: a three-dimensional spatial
region is taken to be some compact volume $V$ with boundary
$\partial V$. The time integration is taken over the interval
$[0, \beta]$ because there is periodicity in Euclidean time.

From the discussion in the preceding section, we now define
the entropy associated with the space-time region to be
$$
S = \frac{1}{8 \pi G} \, \int_{\cal V} \, d^4 x \sqrt{-g} \nabla_i a^i
= \frac{\beta}{8 \pi G} \, \int_{\partial V} \, d^2 x
\sqrt{\sigma} (N n_{\mu} a^{\mu} ).
\eqno(4.1)
$$
The integral is cast into the second form on account of the
previous considerations. The time integration reduces to 
multiplication by $\beta$, and only the spatial components are 
non-zero, so the divergence is three-dimensional over $V$, which
can be transformed into an integral over $\partial V$.
If the boundary $\partial V$ is a horizon, the quantity
$N n_{\mu} a^{\mu}$ will tend to a fixed quantity, namely,  
the constant surface gravity $\kappa$, so the integral 
reduces to an expression for area. Using $\beta \kappa
= 2 \pi$, it is found that $S = {\cal A} / 4G$, where 
${\cal A}$ is the area of the horizon. Similar considerations
apply to each piece of any area element when it acts as a
horizon for some Rindler observer.

The total energy $E$ in the region, acting as a sourse for
gravitational acceleration, is given by the Tolman energy
defined by
$$
E = 2 \int_V \, d^3 x \sqrt{\gamma} N ( T_{ij}
- \frac{1}{2} T \, g_{ij} ) u^i u^j.
\eqno(4.2)
$$
The covariant combination $2 ( T_{ij} - \frac{1}{2} T g_{ij})
u^i u^j$, which reduces to $(\rho + 3 p)$ for an ideal fluid,
is the correct source for gravitational acceleration.
Note that both $S$ and $E$ depend on the congruence of
timelike curves chosen to define them through $u^i$. The free
energy of space-time must have direct geometrical meaning
independent of the congruence of observers used to define
the entropy $S$ and $E$. The energy in (4.2) is not just
$$
U = \int_V \, d^3 x \, \sqrt{\gamma} N ( T_{ij} u^i u^j),
\eqno(4.3)
$$
based on $\rho = T_{ij} u^i u^j$, but the integral of
$( \rho + 3 p)$. The free energy needs to be defined as
$F \equiv U - T \, S$, since pressure, which is an
independent thermodynamic variable, should not
appear in the free energy. This gives,
$$
\beta F = \beta U - S =-S + \beta \int_V \, d^3 x 
\sqrt{\gamma} N ( T_{ij} u^i u^j )
=- S + \int_{\cal V} \, d^4 x \sqrt{-g} T_{ij} u^i u^j.
\eqno(4.4)
$$
Using the expression for the entropy (4.1), 
$R =- 8 \pi GT$ and (3.3)-(3.4), it is found that
$$
\beta F =- \frac{1}{8 \pi G}
\int_{\cal V} \, d^4 x \sqrt{-g} \nabla_i a^i 
+ \int_{\cal V} \, d^4 x \sqrt{-g} T_{ij} u^i u^j
$$
$$
= \int_{\cal V} \, d^4 x \sqrt{-g} ( - \frac{1}{8 \pi G}
\nabla_i a^i + T_{ij} u^i u^j )
= \frac{1}{2} \int_{\cal V} \, d^4 x \sqrt{-g} T \, g_{ij} u^i u^j
=- \frac{1}{8 \pi G} \int_{\cal V} \, d^4 x \sqrt{-g}
R g_{ij} u^i u^j
$$
$$
= \frac{1}{8 \pi G} \, \int_{\cal V} \, d^4 x \sqrt{-g} \, R.
\eqno(4.5)
$$
The equation (4.5) is just the Einstein-Hilbert action. 
The equations of motion are obtained by minimizing the action.
They can be equivalently thought of as arising by minimizing the
macroscopic free energy.

\begin{center}
{\bf 5. Horizons Applied to Other Physical Situations and Conclusions.}
\end{center}

It is clear that the presence of horizons which appear
so naturally in the case of black-holes leads to a link
between entropy and the Einstein-Hilbert action.
t'Hooft's horizon algebra posits that ingoing and
outcoming fields with respect to an event horizon are
projected onto the horizon and determined there.
An event horizon contains all field theoretic information
of a black hole.

What can be said with regard to other systems when there
is an acceleration involved, or when the equivalence
principle can be applied. In particular, the state of
motion of the measuring device can affect whether or
not particles are observed to be present {\bf [5,15]}.
The question as to which set of modes furnishes the best
description of a physical vacuum is not easy to answer.
For example, a free-falling detector will not always
register the same particle density as a non-inertial,
accelerating detector. One of the main reasons for 
the vagueness of the particle concept is in its global 
or extended nature. The modes are defined on the whole 
of space-time, and of course, when there are particles
in large enough numbers, there is the possibility that
thermodynamic variables can be assigned to them, such as
an entropy or free energy, as one would do with any gas.
Similarly, the absence of particles signifies the
absence of properties such as entropy as well.
Perhaps the nature of horizons is to limit the extended 
nature when considering the particle concept. Another 
physical example is the system which contains a
charge fixed with respect to a stationary observer 
which is passed by an observer who occupies an
accelerating frame of reference. The accelerating observer
would presume the charge is accelerating and should 
appear to radiate. On the other hand,
the stationary observer would observe the charge at rest
and not radiating.

Perhaps the idea of the physics of horizons and the
holographic principle has a greater generality which
goes beyond the study of black holes where the idea
of a horizon has a very prominent role. 
This is inevitable since a horizon appears automatically
during formation due to the enormous concentration of mass. 
In fact, it is well known that
boundaries appear in other related relativistic contexts as well.
If a timelike curve $x^a (\tau)$ in the space-time is
considered parametrized by the proper time $\tau$ of the
clock moving along that curve, the union of past light
cones $\{ C ( \tau), - \infty \leq \tau \leq \infty \}$
determines whether an observer on $x^a$ can obtain 
information from all events in the space-time or not. If
there is a nontrivial boundary, there will be regions 
in the space-time from which this observer cannot receive
signals, and a family of such curves is usually called 
a congruence. This could be used to define a type of horizon.
Acceleration itself could be a determining factor much
in the way mass determines a horizon in the black hole context.
That is to say, perhaps it is possible for a sufficiently
large acceleration to twist space-time locally in the
vicinity of the accelerated mass that
it can be said a horizon exists at least locally and has a
physical impact. Much of this could 
depend on how far one would like to push the view that
space-time can be regarded as a membrane which can be
distorted, of course.

To see how this might be approached, let us return to
the identity connecting the $\Gamma^2$ Lagrangian
${\cal L}_{bulk}$ and the Einstein-Hilbert Lagrangian
${\cal L}_{grav}$. 
This relation is a purely differential
geometric identity, and can be written in general as
$$
{\cal L}_{grav} = {\cal L}_{bulk} - \nabla_c
[ g_{ab} \frac{\partial {\cal L}_{bulk}}{\partial ( \partial_c g_{ab})}],
\quad
{\cal L}_{bulk} = {\cal L}_{grav} - \nabla_c 
[ \Gamma^j_{ab} \frac{\partial {\cal L}_{grav}}
{\partial ( \partial_c \Gamma^j_{ab} )} ].
$$
These show that the really important degrees of freedom
in gravity are the surface degrees of freedom. It is
quite plausible that   other situations can be described 
by equations similar to these.

\begin{center}
{\bf References.}
\end{center}

\noindent
$[1]$ J. D. Bekenstein, Phys. Rev. {\bf D7}, 2333-2346, (1973). \\
$[2]$ J. D. Bekenstein, Phys. Rev. {\bf D9}, 3292-3300, (1974). \\
$[3]$ S. W. Hawking, Commun. Math. Physics, {\bf 43}, 199-220, (1975).  \\
$[4]$ S. A. Fulling, Phys. Rev. {\bf D7}, 2850-2862, (1973).  \\
$[5]$ S. A. Fulling, Aspects of Quantum Field Theory in Curved Spacetime,
Cambridge University Press, Cambridge, (1989).  \\
$[6]$ R. M. Wald, General Relativity, The University of Chicago Press,
Chicago, (1984).  \\
$[7]$ W. Rindler, Relativity-Special, General and Cosmological, 2nd Ed.,
Oxford Univ. Press (2006).   \\
$[8]$ T. Padmanabhan, Physics Reports, {\bf 380}, 235-320, (2003).  \\
$[9]$ T. Padmanabhan, Physics Reports, {\bf 406}, 49-125, (2005).   \\
$[10]$ A. V\'azquez, H. Quevedo, A. S\'anchez, J. Geom. and Physics, {\bf 60},
1942-1949, (2010).  \\
$[11]$ T. Padmanabhan, Astro. and Space Science, {\bf 285}, 407-417, (2003). \\
$[12]$ T. Padmanabhan, Class. Quantum Grav., {\bf 19}, 3551-3566, (2002).  \\
$[13]$ T. Padmanabhan, Class. Quantum Grav., {\bf 19}, 5387-5408, (2002).  \\
$[14]$ M. Carmeli, Classical Fields, General Relativity and Gauge Theory,
World Scientific, Singapore, (2001).  \\
$[15]$ N. D. Birrell and P. C. W. Davies, Quantum Fields in Curved Space,
Cambridge University Press, (1982).   \\
\end{document}